\documentclass[12pt]{iopart}

\usepackage{graphicx,graphics,epsfig,subfigure}
\usepackage{gensymb}
\usepackage{cite}
\usepackage[matrix,frame,arrow]{xypic}
\usepackage[pdfstartview=FitH]{hyperref}

\usepackage[pdftex]{color}
\usepackage{setspace}

\newcommand{\RNum}[1]{\uppercase\expandafter{\romannumeral #1\relax}}

\begin{document}

\title{Exploring Topological Phase Transition via Quantum Walk in Coherent State Space }
\author{Zi-Yong Ge $^{1,2}$ and Heng~Fan$^{1,2,3}$}

\address{$^1$~Institute of Physics, Chinese Academy of Sciences, Beijing 100190, China}
\address{$^2$~School of Physical Sciences, University of Chinese Academy of Sciences, Beijing 100190, China}
\address{$^3$~CAS Central for Excellence in Topological Quantum Computation, University of Chinese Academy of Sciences, Beijing 100190, China}

\ead{hfan@iphy.ac.cn}

%\keywords{Quantum walks, coherent space, topological quantum phase transition, circuit quantum electrodynamics}

\vspace{10pt}

\begin{abstract}
The quantum walk is a dynamical protocol which describes the motion of spinful particles on a lattice. Also, it has been demonstrated to be a powerful platform to explore topological quantum matter. Recently, the quantum walk in coherent state space has been proposed theoretically and realized experimentally. However, due to the inherent characteristics of coherent states, 
it is challenging to control the number of photons when we need the coherent space to be a nearly orthogonal space in practice. Here, we demonstrate that the nonorthogonality of coherent sates, on the one hand can be cancelled by multiple measurement, on the other hand, it is useful resource to characterize the nature of the system. Thus the number of photons of the system is controllable. We first present a feasible scheme to measure the wave function of quantum walks. Then we show that the expected number of photons of the coherent space is good observable to represent topological properties of the system, which reflected the advantage of coherent state space quantum walks. In addition, we propose an experimental protocol in a circuit quantum electrodynamics architecture, where a superconducting qubit is a coin while the cavity mode is used for quantum walk.	
\end{abstract}

\section{Introduction}

Quantum walks, extended from classical random walks, are designed as the protocol which describes the dynamics
of single spinful particle on a lattice \cite{Aharonov,Kempe,Venegas}. Topological order plays a significant role in modern condensed matter physics, since it is beyond Landau symmetry-breaking theory \cite{Su,Thouless,Haldane1,Haldane2,Wen1,Chen,Tsui,Laughlin,Kitaev1,Chiu,Kane1,Kane2,Hasan,Qi,Read,Kitaev2}. Recently, Kitagawa \textit{et al}.\cite{Kitagawa1} found that quantum walk is a powerful platform to explore the topological phases, especially symmetry protected topological order \cite{Chiu}. Subsequently, a lot of experiments have observed such topological features, including edge-bulk correspondence, topological invariants and topological phase transitions \cite{Kitagawa2,Flurin,Cardano1,Cardano2,Zhan}. In addition, the non-Hermitian quantum walks are also proposed and realized experimentally \cite{Rudner,Zhan},
which can be used to study topological phase. There are many methods or platforms to implement quantum walks, including optical system \cite{Kitagawa2,Cardano1,Zhan,Cardano2}, trapped ions \cite{Schmitz,Zahringer,Travaglione} and superconductor quantum circuits  \cite{Flurin,Xue,Ramasesh}.

Generally, quantum walks are studied in real space. As the development of cavity and circuit quantum electrodynamics (QED)  \cite{Haroche,Chang, Sanders,Koch}, the quantum walk in coherent state space (CSS) was proposed theoretically and realized experimentally \cite{Flurin,Schmitz,Xue,Ramasesh}. The most advantage of CSS quantum walk is that it is extensible, so that it is convenience to realize high dimensional quantum walks and increase the walk steps. Nevertheless, since arbitrary two coherent states are not orthogonal to each other, when we just concern the corresponding information of real space quantum walks, the experimental error is large. To overcome this difficulty, the practice is to let the distance of two adjacent coherent states be large enough, so that this coherent space can be considered as a similar orthogonal space. However, it will lead the number of photons uncontrollable, which may arise new problems, for instance, undesired nonlinear effects. All in all, it is still a challenge to control the photon number and experimental error of the system at the same time.

\begin{figure}
	\centering
	\vspace{0.2cm}
	\includegraphics[width=0.35\textwidth]{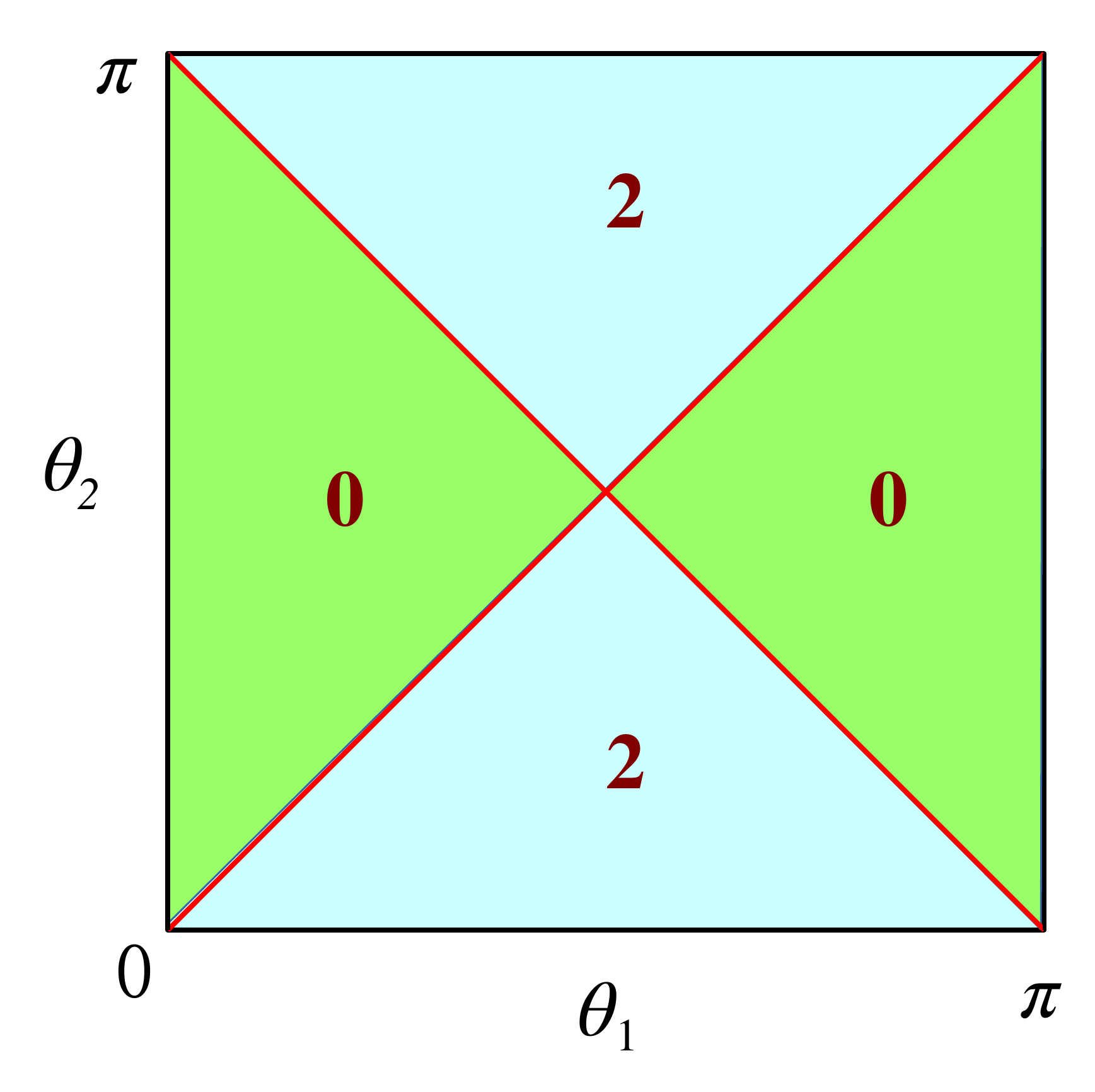}	
	\caption{(Color online) The phase diagram of the Hamiltonian. Here, parameters in (\ref{heff}) are as, $\theta_1,\theta_2 \in [0,\pi]$. The critical points satisfy $|\frac{\tan\theta_1}{\tan\theta_2}|=1$.  When $|\frac{\tan\theta_1}{\tan\theta_2}|>1$, the winding number is 2, i.e., topological nontrivial.	When $|\frac{\tan\theta_1}{\tan\theta_2}|<1$, it is topological trivial phase.}
	\label{pd}
\end{figure}

Motivated by above consideration, in this paper, we will explore topological phase transition in CSS quantum walks, 
and present several schemes to guarantee that the photon number and experimental error can both be controlled simultaneously. 
We find that, the inherent non-orthogonality of coherent space, on the one hand, can be cancelled by  multiple measurement,
on the other hand, it is also useful resource to represent the nature of the system. Therefor, in this framework, the number of photons can be absolutely controllable.
Furthermore, we demonstrate that the average number of photons can actually be enough to characterize topological phase transition, which can be measured easily in experiments.
Additionally, we design an experiment in superconducting quantum circuit to realize CSS quantum walks, where the superconducting qubit couples to a microcavity with large detuning \cite{Schmitz,Xue,Ramasesh}..

The paper are organized as follows. In section \RNum{2}, we review the split-step quantum walks in real space and define the first and second moments to characterize topological phase transition. In section \RNum{3}, we start to study CSS quantum walks and show that how to use the intrinsic resource of coherent states. We first present a realizable scheme to measure corresponding wave functions of real space quantum walks for CSS quantum walks. Then we show that the average number of photons is a good observable to represent the topological properties of systems. In section \RNum{4}, an explicit experiment based on circuit QED is proposed.

\section{Split-step quantum walks}
Generally, a quantum walk contains a coin and a walker which represent the spin sate and position state respectively. Here we introduce a special quantum walk, that is, split-step quantum walk (SSQW), which alternates two coin tosses between two spin-dependent translations.
In real space, its unitary operator of each step can be written as
\begin{eqnarray}\label{uw}
\hat{U}_W(\theta_1,\theta_2) = \hat{T}_{\uparrow\downarrow}\hat{R}_x(2\theta_2)\hat{T}_{\uparrow\downarrow}\hat{R}_x(2\theta_1),
\end{eqnarray}
where
\begin{eqnarray}\label{tud}
\hat{T}_{\uparrow\downarrow} = \sum_x(|x\!+\!1\rangle\langle x|\otimes|\!\!\uparrow\rangle\langle\uparrow\!\!|+|x\!-\!1\rangle\langle x|\otimes|\!\!\downarrow\rangle\langle\downarrow\!\!|)
\end{eqnarray}
is spin-dependent translation ($x$ is integer) and $\hat{R}_x(2\theta) = e^{-i\theta\hat{\sigma}_x}$ is spin rotation. Using Fourier transformation
\begin{eqnarray}\label{ft}
|k\rangle =\frac{1}{ \sqrt{2\pi}}\sum_xe^{-ikx}|x\rangle,
\end{eqnarray}
the spin-dependent operator can be diagonalized as
\begin{eqnarray}\label{dig}
\hat{T}_{\uparrow\downarrow} = e^{ik\sigma_z}\otimes|k\rangle\langle k|.
\end{eqnarray}
Since
\begin{eqnarray}\label{hef}
\hat{U}_W(\theta_1,\theta_2) = e^{-iH_{\text{eff}}(\theta_1,\theta_2)},
\end{eqnarray}
we can obtain the effective Hamiltonian \cite{Kitagawa1} as
\begin{eqnarray}\label{heff} 
H_{\text{eff}}(\theta_1,\theta_2)  =  \sum_k \epsilon_{\theta_1,\theta_2}(k)\bi{n}_{\theta_1,\theta_2}(k)\cdot\bi{\sigma}\otimes|k\rangle\langle k|. 
\end{eqnarray}
Here, $\epsilon_{\theta_1,\theta_2}(k)$ characterizes the band structure while $\bi{n}_{\theta_1,\theta_2}(k)$ corresponds to the eigenstates of single particle in terms of lattice momentum $k$. They have the form
\begin{eqnarray}\label{en}  \nonumber
\cos&& (\epsilon_{\theta_1,\theta_2}(k))= \cos(2k)\cos\theta_1\cos\theta_2+\sin\theta_1\sin\theta_2 \\ [0.2cm] \nonumber
&&n_x(k) =\frac{\cos(2k)\cos\theta_1\sin\theta_2+\sin\theta_1\cos\theta_2 }{\sin(\epsilon_{\theta_1,\theta_2}(k))} \\[0.2cm]  \nonumber
&&n_y(k) =-\frac{\sin(2k)\cos\theta_2\sin\theta_1 }{\sin(\epsilon_{\theta_1,\theta_2}(k))} \\ [0.2cm]
&&n_z(k) =-\frac{\sin(2k)\cos\theta_2\cos\theta_1 }{\sin(\epsilon_{\theta_1,\theta_2}(k))} .
\end{eqnarray}
We can find that when $|\frac{\tan\theta_1}{\tan\theta_2}|=1$, $\epsilon_{\theta_1,\theta_2}(k)=0$, i.e., the  gap closes, which indicates topological phase transition. Furthermore, $H_{\text{eff}}(\theta_1,\theta_2)$ has time reversal ($\mathcal{T}$), particle-hole ($\mathcal{P}$) and chiral symmetries ($\Gamma$), satisfying $\mathcal{T}^2 = \mathcal{P}^2= 1$ and $\Gamma = e^{-i\pi\bi{A}\cdot\bi{\sigma}/2}$ where $\bi{A} =  (0,\cos\theta_1,-\sin\theta_1)$. Therefor, this Hamiltonian is a typical SPT order. Using tenfold classification\cite{Chiu,Kitaev3}, it belongs to class BD\RNum{1}, which means that its topological invariant is winding number defined as
\begin{eqnarray}\label{wn}
\gamma = \frac{1}{2\pi}\!\int_{-\pi}^{\pi}\text{d}k\big(\bi{n}\times\frac{\partial \bi{n}}{\partial k}\big)\cdot \bi{A}.
\end{eqnarray}
When $\gamma\!\!\neq\!\! 0$, it is topological nontrivial, and the phase diagram is shown in Fig.\ref{pd}. Without loss of generality, in the following discussion, unless otherwise specified, $\theta_1$ is fixed to $\pi/4$.  The critical points are at $\theta_2=\pi/4$ and $3\pi/4$.

\begin{figure}[b]
	\centering
	\subfigure{
		\includegraphics[width=0.7\textwidth]{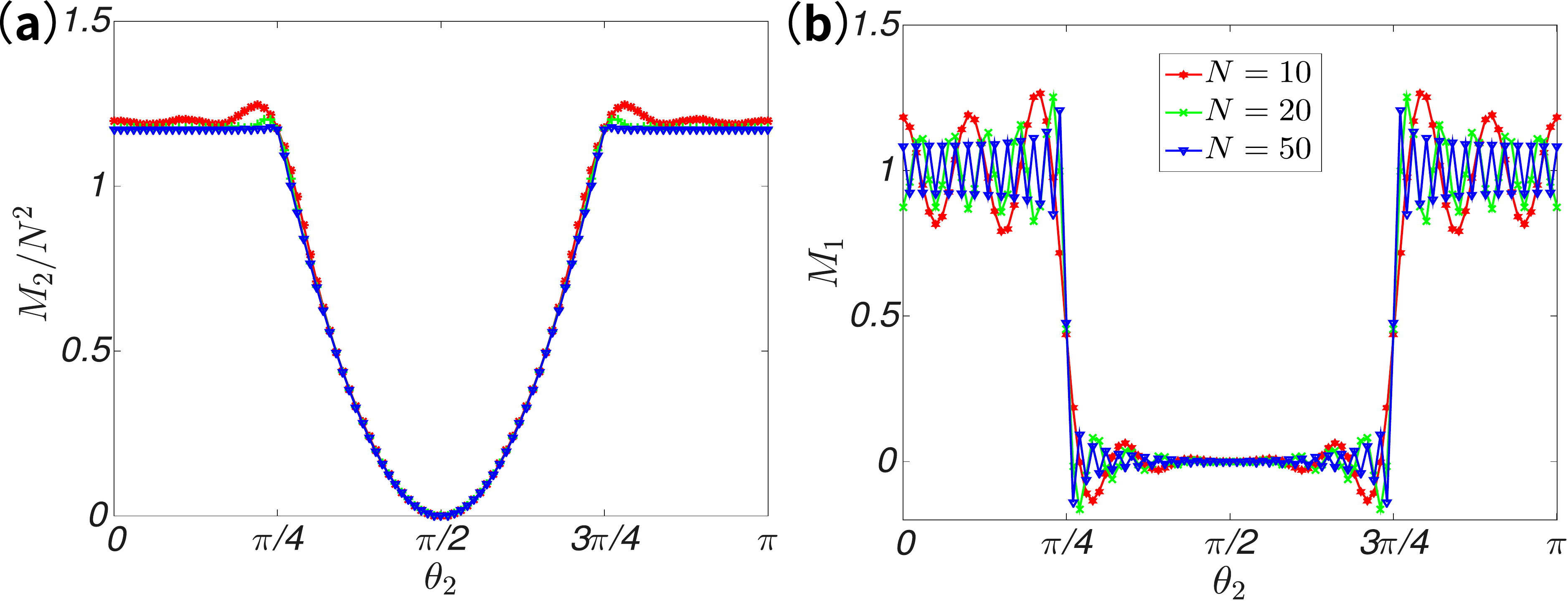}	}	
	\caption{(Color online)  The first and second moments for real-space quantum walks. The results for walk step $N = 10,20,50$ are shown respectively. (\textbf{a}) The initial state is $|0\rangle\otimes|\!\!\uparrow\rangle$. In topological nontrivial phase, $M_1$ keeps constant when varying $\theta_2$. With the increase of walk steps, the nonanalyticity at critical point is more and more distinct. (\textbf{b}) The initial state is $|0\rangle\otimes((\sqrt{2}+1)i|\!\uparrow\rangle+|\!\downarrow\rangle)/\sqrt{2\sqrt{2}+4}$, of which the coin state is the eigenstate of $\Gamma$ and the eigenvalue is $1$. The equilibrium position tends to $\gamma/2=1$ in topological topological nontrivial phase, while in topological trivial phase, it tends to $0$. }
	\label{m1m2}
\end{figure}

To describe phase transition, we need the first and second moments. According to Refs. \cite{Cardano1,Cardano2}, the \textit{j}-th moments are defined as
\begin{eqnarray}\label{jth}
M_j(N) = \sum_x x^j P_N(x),
\end{eqnarray}
where $x$ represents the walker position while $P_N(x)$ is the relevant probability distribution after $N$ steps.
Firstly, let us calculate $M_2$. Considering $N\!\rightarrow\!\infty$, we have
\begin{eqnarray}\label{m2n}
M_2(N)/N^2  = L(\theta_1,\theta_2) + O(1/N^2),
\end{eqnarray}
and
\begin{eqnarray}\label{L12}
L(\theta_1,\theta_2) =\int_{-\pi}^{\pi}\frac{\text{d}k}{2\pi}[V_{k}(\theta_1,\theta_2)]^2 .
\end{eqnarray}
Here, $V_{\theta_1,\theta_2}(k) := \text{d}\epsilon_{\theta_1,\theta_2}(k)/\text{d}k$, is the associated group velocity. According to the residue theorem, we know $L(\theta_1,\theta_2)$ is nonanalytical at critical points, since gappless points are the poles of $V_{\theta_1,\theta_2}(k)$. So $M_2$ can characterize the phase transition just like the order parameter. Fig. \ref{m1m2}(a) shows the relation between $M_2$ and $\theta_2$ when $N$ choose different values.

As for the first moment $M_1$, similar to Ref. \cite{Cardano2}, we have,
\begin{eqnarray}\label{m1} 
M_1(N)= \langle\Gamma_\perp\rangle_{\psi_0} [L(t)+S(t)]-\langle\Gamma\rangle_{\psi_0} S_{\Gamma}(t).
\end{eqnarray}
Here, $|\psi_0\rangle$ is the initial state and $\bi{s} = \langle\bi{\sigma}\rangle_{\psi_0}$, and
\begin{eqnarray}\label{sgmat} 
S_{\Gamma}(t)\!  = \!\frac{\gamma}{2} \!-\!\frac{1}{2} \int_{-\pi}^{\pi}\!\frac{\text{d}k}{2\pi}\cos(2N\epsilon)\big(\bi{n}\!\times\!\frac{\partial\bi{n}}{\partial k}\big)\big]\cdot \!\bi{A}, 
\end{eqnarray}
where $\gamma$ is the winding number. If the initial spin polarization direction is orthogonal to $\bi{A}$, i.e., $\langle\Gamma_\perp\rangle_{\psi_0}=0$, then
\begin{eqnarray}\label{m1n} 
M_1(N)\! = \! \pm \!\frac{\gamma}{2}\! \mp\!\frac{1}{2}\! \int_{-\pi}^{\pi}\!\frac{\text{d}k}{2\pi}\cos(2N\epsilon)\big(\bi{n}\!\times\!\frac{\partial\bi{n}}{\partial k}\big)\big]\!\cdot\! \bi{A}. 
\end{eqnarray}
So when $N\!\rightarrow\!\infty$, the second term of Eq.(\ref{m1n}) quickly converges to zero. It can thus show obvious topological phase transition, and see Fig. \ref{m1m2}(b).

\section{Coherent state space quantum walks}
In this section, we discuss the walker walks in coherent space, that is, CSS quantum walks, by replacing real-space state $|x\rangle$ with coherent state
\begin{eqnarray}\label{xa}
|x\alpha\rangle=e^{-|x\alpha|^2/2}\sum_{n}x^n\alpha^n/\sqrt{n!}|n\rangle.
\end{eqnarray}
Hence, for SSQW, the spin-dependent-translation operator, $\hat{T}_{\uparrow\downarrow}(\alpha)$, satisfies
\begin{eqnarray}\label{tud2}
\hat{T}_{\uparrow\downarrow}(\alpha)|x\alpha,\uparrow\downarrow\rangle=|(x\pm1)\alpha,\uparrow\downarrow\rangle.
\end{eqnarray}
Thus, the final state has the form of $|\psi_f\rangle = \sum_{x,\sigma} A_{x,\sigma}|x\alpha,\sigma\rangle$, where $\sigma$ is the spin index. Obviously, CSS quantum walks contain the whole information of real space quantum walks.

Now we consider how to extract $M_1$ and $M_2$ to represent topological properties. The most immediate approach is finding wave function, i.e., $P_N(x)$, then calculate them according to Eq.(\ref{jth}). In fact, for CSS quantum walks, we can also obtain $M_1$ and $M_2$ by means of
measuring expected photon number.
In the following contents of this section, we will show more details of these two methods.

\begin{figure*}
	\centering
	\includegraphics[width=0.9\textwidth]{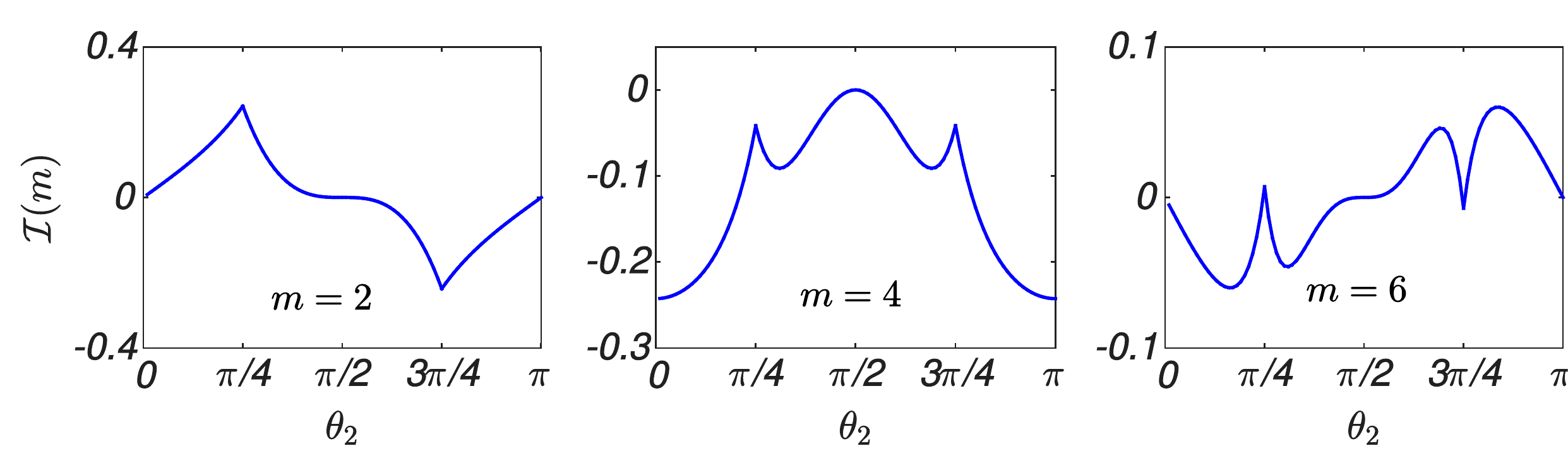}		
	\caption{(Color online) ( The function between $\theta_2$ and $\mathcal{I}(m=\text{even})$ with $m = 2,4,6$,  respectively. There are distinct pinnacles at critical points, which are nonanalytical and represent phase transition. }
	\label{im}
\end{figure*}

\subsection{Finding wave function}
As mentioned, the number of photons is expected to be controllable, so $|\alpha|$ can not be so large. Thus, arbitrary two coherent states with short distance are not orthogonal, in such case, $P_N(x)$ cannot be measured directly. However, we could cancel the contribution of this nonorthogonality. Consider the projective operator, defined as $\hat{\Pi}(x\alpha)\equiv|x\alpha\rangle\langle x\alpha|$, which measures the probability of $|x\alpha\rangle$, that is,
\begin{eqnarray}\label{px} \nonumber
P(x\alpha) &&= \!\langle\psi_f|\hat{\Pi}(x\alpha)|\psi_f\rangle = \!\!\!\sum_{x',x''}\!\! A_{x'}A_{x''}^*\langle x\alpha|x'\alpha\rangle\langle x''\alpha|x\alpha\rangle \\ [0.1cm]  \nonumber
&&= |A_x|^2 + e^{-|\alpha|^2/2}(A_{x+1}^*A_{x}+A_{x-1}^*A_{x} + c.c) \\[0.2cm] \nonumber
&& + e^{-|\alpha|^2}(|A_{x+1}|^2+|A_{x-1}|^2 + 2\text{Re} (A_{x+1}^*A_{x-1})) \\ [0.2cm] 
&&+ e^{-2|\alpha|^2}(A_{x+2}^*A_{x}+A_{x-2}^*A_{x} + c.c) + ...
\end{eqnarray}
\noindent Here, $c.c$ is complex conjugation, and for convenience, we neglect the spin index $\sigma$, which do not affect the results. Let $t \equiv e^{-|\alpha|^2/2}$, and Eq. (\ref{px}) has the form
\begin{eqnarray}\label{px2}
P(x\alpha)= c_0+c_1t+c_2t^2 + c_4 t^4 +....
\end{eqnarray}
Obviously, here, $c_i$ is \textit{t}-independent (or $\alpha$-independent), and $c_0$ is what we need to find, i.e., $P_N(x)$. Now consider several CSS quantum walks with different $\alpha$s, and we measure their $P(x\alpha)s$ respectively. Then we can fix $c_0$ by means of polynomial fitting.

Generally, the error of polynomial fitting may be too large to lead the failure of experiments. Nevertheless, for quantum walks, on the one hand, $P(x\alpha)$ converges quickly, so that the high order can be cut. One the other hand, due to some inherent symmetries, the terms which contributes can be reduced further. For instance, consider SSQW, since each step contains two translations, after several walks, the odd sites vanish, i.e., $A_{x=\text{odd}}=0$. Therefor, in principle, the practical complexity and error are both acceptable.

\subsection{Expected number of photons}
When the step is so large, measuring the wave function will not be realizable, and we thus need to find other observables to represent the characters of the system, for instance, topological properties. Now we show that, for CSS quantum walks, the expected photon number of the systems is a good observable to represent $M_1$ and $M_2$.

The expected photon number of the CSS quantum walks, $N_w$, can be obtained as
\begin{eqnarray}\label{adagea} \nonumber
N_w&&=\langle \hat{a}^\dagger\hat{a}\rangle  =  \sum_{x} x^2|A_{x}|^2|\alpha|^2 \\ \nonumber
&&  +\sum_{x}x(x+1)A_{x}^*A_{x+1}|\alpha|^2e^{-\alpha^2/2} + c.c\\ 
&& +\sum_{x} x(x+2)A_{x}^*A_{x+2}|\alpha|^2e^{-2\alpha^2} + c.c+ \cdot\cdot\cdot . 
\end{eqnarray}

First of all, if $|\alpha|$ is large enough, only the first term contributes in Eq. (\ref{adagea}). In such case, $N_w=|\alpha|^2 M_2$, so $N_w$ can indeed characterize topological phase transition.
In addition, if there are two CSS quantum walks, of which initial states are $|m\alpha\rangle\otimes|s\rangle$ and $|-m\alpha\rangle\otimes|s\rangle$, respectively, then
\begin{eqnarray}\label{nw12} \nonumber
&&N_{w1}= \sum_{x} |\alpha|^2(x+m)^2P_x \\
&&N_{w2}= \sum_{x} |\alpha|^2(x-m)^2P_x ,
\end{eqnarray}
thus,
\begin{eqnarray}\label{dnw}
\Delta N_w = N_{w2}-N_{w1}=4m|\alpha|^2M_1.
\end{eqnarray}
When the polarization direction of $|s\rangle$ is orthogonal to $\textit{\textbf{A}}$, $\Delta N_w$ is also able to characterize topological phase transition as mentioned in Eq. (\ref{m1n}).

\begin{figure}
	\centering	
	\includegraphics[width=0.8\textwidth]{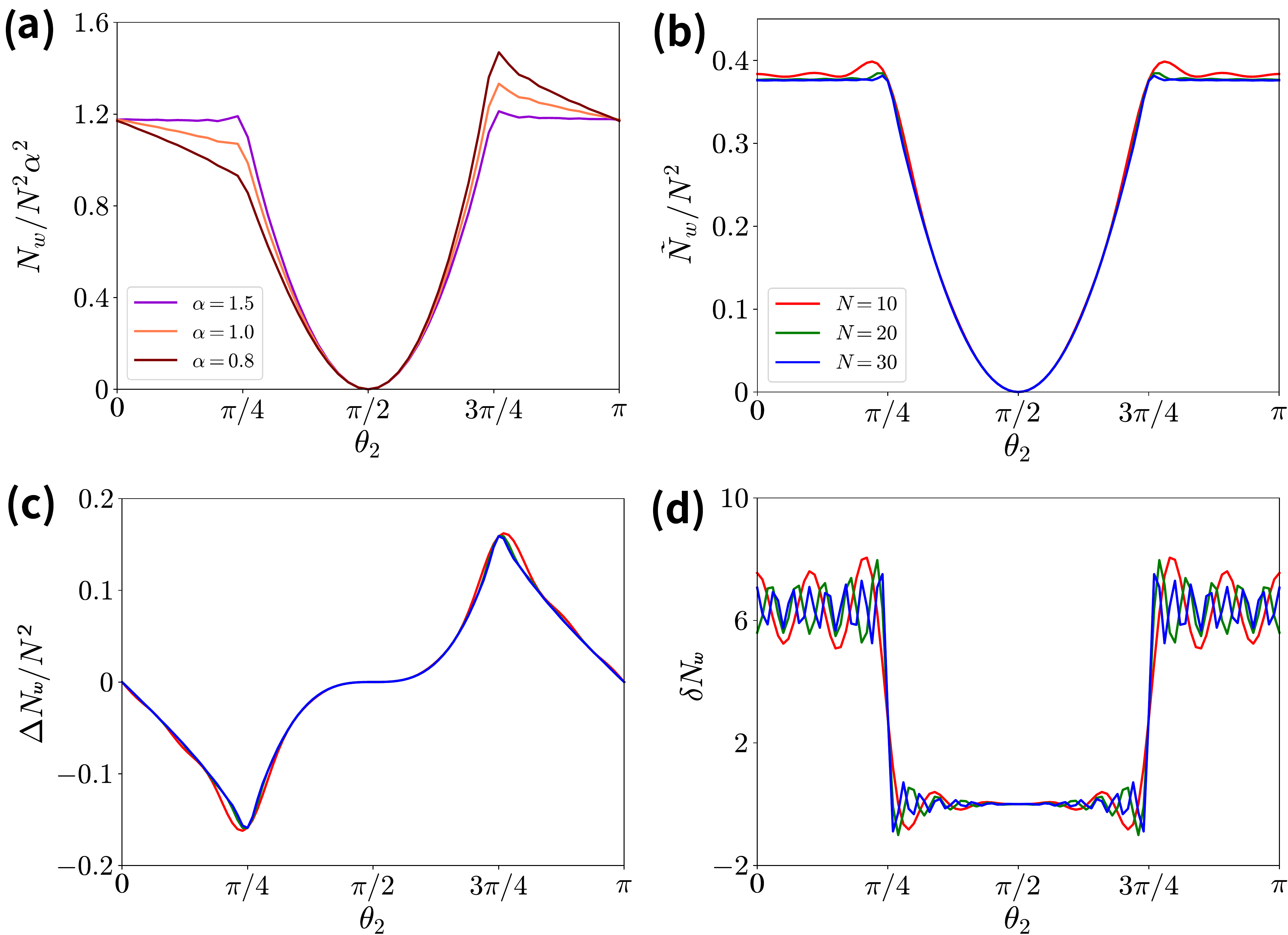}	
	\caption{(Color online) (\textbf{a}) The average number of photons with walk steps $N=20$ and $\alpha=1.5,\ 1.0,\ 0.8$ respectively. When $\alpha=1.5$,  $\mathcal{I}(2)$ has not contributed, so $N_w(\alpha=1.5)$ is proportional to $M_2$. When $\alpha=1.0,\ 0.8$,  $N_w$s deviate from $M_2$, since the higher terms are not able to be neglected, however, at critical points, $N_w$s are also nonanalytical. (\textbf{b})-(\textbf{d}) The average number of photons, when adding $R_z(2\phi)$. we can separate the different terms which contribute $N_w$. Here, $\tilde{N}_w,\ \Delta N_w$ and $\delta N_w$ are displayed with $\alpha=0.4$ and walk steps $N=10,\ 20$ and $30$, respectively. For (a)-(b), the initial state is $|0\rangle\otimes|\!\!\uparrow\rangle$, while for (d), the initial states are $|2\rangle\otimes|s\rangle$ and $|\!-\!2\rangle\otimes|s\rangle$ where $|s\rangle=((\sqrt{2}+1)i|\!\uparrow\rangle+|\!\downarrow\rangle)/\sqrt{2\sqrt{2}+4}$.}
	\label{NW}
\end{figure}

Similarly, we consider the case that $|\alpha|$ is not very large, therefor, for Eq. (\ref{adagea}), the second and third terms (or higher terms) cannot be neglected,
and we denote the $j\!\!+\!\!1$-th term as $\mathcal{T}_j$. Consider another state in terms of real space
$|\psi'\rangle = \sum_{x} A_{x}|x,\rangle$, which can be regarded as the final state of real-space quantum walk by just replace coherent-state basis $|x\alpha\rangle$ with real-space-state basis $|x\rangle$, so
\begin{eqnarray}\label{tj}
\mathcal{T}_j = |\alpha|^2e^{-j^2\alpha^2/2}\langle\psi'|\hat{x}^2(\hat{T}(j)+\hat{T}(-j))|\psi'\rangle. 
\end{eqnarray}
Here, $\hat{T}(j)$ is translation operator and has the form of $e^{-i\hat{k}j}$. Following Eq. (\ref{tj}) to calculate $N_w$, we have
\begin{eqnarray}\label{nw} \nonumber
N_w &&= |\alpha|^2 M_2 - 2|\alpha|^2e^{-\alpha^2/2}\langle \hat{U}_W^{-N}
\partial_k^2 \cos k\hat{U}_W^N\rangle_{\psi_0} \\ [0.2cm]
&&-2|\alpha|^2e^{-2\alpha^2}\langle \hat{U}_W^{-N} \partial_k^2 \cos 2k\hat{U}_W^N\rangle_{\psi_0} +\ \cdot\cdot\cdot.
\end{eqnarray}
Here, the unitary operator $\hat{U}_W$ and initial state $\psi_0$ both correspond to real-space quantum walk. For convenience, let
\begin{eqnarray}\label{im_}
\mathcal{I}(m) \equiv \langle \hat{U}_W^{-N} \partial_k^2 \cos (mk)\hat{U}_W^N\rangle_{\psi_0}/N^2,
\end{eqnarray}
and obviously $\mathcal{I}(0)=M_2/N^2$. Since $A_{x=\text{odd}}=0$, for $N\rightarrow\infty$, $\mathcal{I}(m=\text{odd})=0$, while $\mathcal{I}(m=\text{even})$ are nonanalytical at critical points. So $\mathcal{I}(m)$ are the function of $\theta_2$, and the curves show in Fig. \ref{im}, when $m = 2,4,6$, respectively. When $|\alpha|$ is not so small, i.e.,
$\mathcal{I}(2)$ has no contribution yet, $N_w$ is still proportional to $M_2$. However, if $|\alpha|$ is so small, $N_w$ will deviate from $M_2$, see Fig. \ref{NW}(a).

If we want to reduce further the photon number, we need to deal with the second and fourth terms.
Now, we add a spin rotation operator around $z$-axis after each $\hat{T}_{\uparrow\downarrow}(\alpha)$, so that new unitary operator of one walk step is
\begin{eqnarray}\label{uw_} 
\tilde{U}_W(\theta_1,\theta_2)\!\! =\!\!\hat{R}_z(2\phi) \hat{T}_{\uparrow\downarrow}(\!\alpha\!)\hat{R}_x(\!2\theta_2\!)\hat{R}_z(2\phi)\hat{T}_{\uparrow\downarrow}(\!\alpha\!)\hat{R}_x(\!2\theta_1\!)
\end{eqnarray}
Thus, it can accumulate different phases for different spin directions during each of the walks, that is, the final state of system is $|\psi'_f\rangle = \sum_{x,\sigma} e^{ix\phi}A_x|x\alpha,\sigma\rangle$. In such case, $N_w$ can be obtained as
\begin{eqnarray}\label{nw2} \nonumber
N_w && = |\alpha|^2 M_2 - 2\cos2\phi|\alpha|^2N^2e^{-\alpha^2/2}\mathcal{I}(2)\\  [0.2cm]
&&-2\cos4\phi|\alpha|^2N^2e^{-2\alpha^2}\mathcal{I}(4)+\ \cdot\cdot\cdot.
\end{eqnarray}
Now, the question is how to choose $\phi$. Consider two walks, of which $\phi$ are $\pi/8$ and $3\pi/8$, respectively, and their initial states are both $|0\rangle\!\otimes\!|s\rangle$.
By Calculating the average number of photons of these two walks, we have
\begin{eqnarray}\label{n_w} \nonumber
&&  N_w(\!\phi\!\!=\!\!\pi/8\!)\!\!=\!\!|\alpha|^2\! M_2\!-\sqrt{2}N^2e^{-\alpha^2\!/2}\mathcal{I}(\!2\!)\\[0.3cm] 
&& N_w(\!\phi\!\!=\!\!3\pi\!/8)\!=|\alpha|^2 M_2+\sqrt{2}N^2e^{-\alpha^2/2}\mathcal{I}(2). 
\end{eqnarray}
We define the summation and difference of two $N_w$ as
\begin{eqnarray}\label{tdn} \nonumber
\tilde{N}_w \!=N_w(\phi=\!\pi/8)\!+\!N_w(\phi\!=\!3\pi/8)\!= \!2|\alpha|^2 M_2\\[0.3cm] \nonumber
\Delta N_w/N^2=N_w(\phi=\pi/8)-N_w(\phi=3\pi/8)  
=-2\sqrt{2}|\alpha|^2e^{-\alpha^2/2}\mathcal{I}(2). \\
\end{eqnarray}
So $\tilde{N}_w$ has the information of $M_2$ while $\Delta N_w$ has the information of $\mathcal{I}(2)$, which means that they can both characterize quantum phase transition [Fig. \ref{NW}(b-c)].

In addition, the information of $M_1$ can also be extracted via $N_w$. We consider two groups of walks, and each group has two walks with same initial state but different $\phi$s. For the first group, the initial state is $|m\alpha\rangle\otimes|s\rangle$ while $\phi=\pi/8$, $3\pi/8$ respectively, so that
\begin{eqnarray}\label{nwt1}
\tilde{N}_{w1} = 2|\alpha|^2( M_2+2mM_1+m^2 ).
\end{eqnarray}
As for the second, we just change the initial sate into $|\!-\!m\alpha\rangle\otimes|s\rangle$, therefor,
\begin{eqnarray}\label{nwt2}
\tilde{N}_{w2} = 2|\alpha|^2( M_2-2mM_1+m^2 ).
\end{eqnarray}
The difference between $\tilde{N}_{w1}$ and $\tilde{N}_{w2}$ is
\begin{eqnarray}\label{Dnw}
\delta N_w =\tilde{N}_{w1}-\tilde{N}_{w2}=8m|\alpha|^2 M_1.
\end{eqnarray}
As mentioned in Eq. (\ref{m1}), when $|s\rangle$ is the eigenstate of $\Gamma$, $\delta N_w$ can characterize the topological invariant of the quantum walk [Fig. \ref{NW}(d)].

\section{Experimental scheme}
We now propose an experimental scheme to realize CSS quantum walk and detect topological phase transition in superconducting circuit. In previous works \cite{Flurin,Xue,Ramasesh}, using circuit QED to implement quantum walks in phase space of cavity mode had been proposed and realized experimentally. Here, we let walker walks in coherent space of cavity mode in a line while the superconducting qubit represents the internal spin of the walker.

The spin rotation operator $R_x(\theta)$ and $R_z(\phi)$ can be performed by microwave driving with high fidelity.
The key problem here is how to implement the spin-dependent translation $\hat{T}_{\uparrow\downarrow}(\alpha)$.
Analogous to Refs. \cite{Flurin,Xue,Ramasesh}, we consider that the superconducting qubit and the cavity satisfy dispersive coupling,
of which free evolution $\hat{U}(t) = e^{-igt\hat{a}^{\dagger}\hat{a}\hat{\sigma}_z}$ in rotating coordinate, where $g$ is coupling strength,
$\hat{a}^{\dagger}$($\hat{a}$) is creation (annihilation) operator of photons.

\begin{figure}
	\centering
	\subfigure{
		\includegraphics[width=0.6\textwidth]{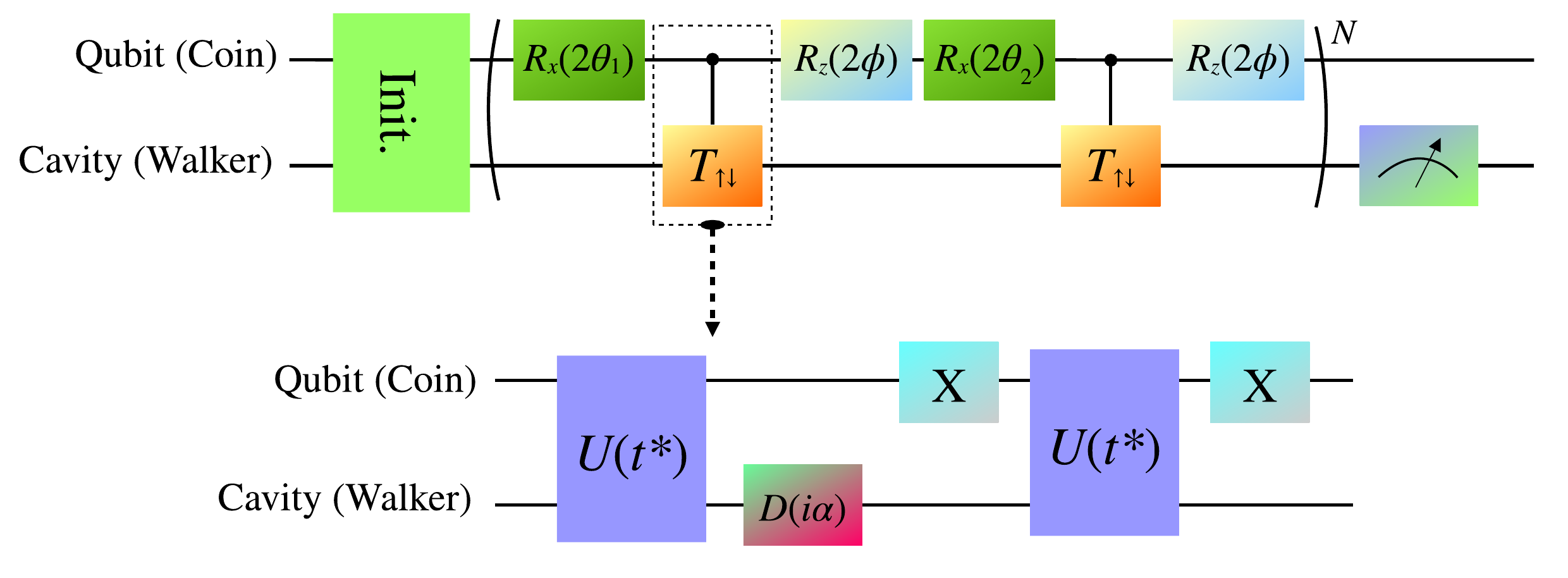}	}	
	\caption{(Color online) The diagram of quantum circuit corresponding to $N$-step CSS quantum walk in c-QED. Here, the superconducting qubit is a coin while the cavity is a walker. The top one is the total circuit, including initialization, quantum walk and measurement. The bottom one is the detail of $\hat{T}_{\uparrow\downarrow}(\alpha)$.}
	\label{cqed}
\end{figure}

Let $t = t^*$, such that $gt^*=\pi/2$, and thus, $\hat{U}(t^*)|\alpha,\uparrow\downarrow\rangle=|\pm i\alpha,\uparrow\downarrow\rangle$.
In addition, we need displacement operator $\hat{D}(\alpha) = e^{\alpha\hat{a}^{\dagger}-\alpha^*\hat{a}}$, which can also be performed experimentally and satisfies $\hat{D}(\alpha)|\beta\rangle = e^{(\alpha\beta^*-\alpha^*\beta)/2}|\alpha+\beta\rangle$. Thus, the spin-dependent translation operator can be implemented by such operation sequence: $\hat{U}(t^*)\rightarrow\hat{D}(i\alpha)\rightarrow\hat{U}(3t^*)$, that is, $\hat{T}_{\uparrow\downarrow}(\alpha)=\hat{U}(3t^*)\hat{D}(i\alpha)\hat{U}(t^*)$.
Here, the time of $t^*$ might be too long, we can use two $\pi$ pulses (\textit{X} gate) and $U(t^*)$ to replace $\hat{U}(3t^*)$, that is, $U(3t^*) = \hat{\sigma}_x\cdot\hat{U}(t^*)\cdot\hat{\sigma}_x$. Furthermore, this sequence has another advantage that it can realize dynamical-decoupling effect (or spin echo) which can prolong coherent time of qubit \cite{Viola}.

For measurement, we just need to detect the final average number of photons in cavity after several walk steps.
In fact, we even do not need to detect the absolute number of photons, while the relative optical intensity is enough.
The explicit quantum circuit is presented in Fig. \ref{cqed}.

\section{Conclusions}
In summary, we have demonstrated that CSS quantum walks can detect the topological phase transition by measuring the wave function expected photons number of the system, and the resource of coherent states is utilized. Meanwhile, we can control both the number of photons and experimental error when walk step is too large.
In this paper, the case of closed system is investigated, and we look forward to extending this idea to dissipated systems or non-Hermitian systems.
Furthermore, high dimensional CSS quantum walks are also worth exploring.

\section*{Acknowledgement}
We thank Zhan Wang, Yu-Ran Zhang and Kai Xu for useful discussion. This work was supported by the State Key Development Program for Basic Research of China
(Grant Nos. 2016YFA0302104£¬2016YFA0300600), the National Natural Science Foundation of China (Grant Nos. 91536108, 11774406), and Strategic Priority Research Program of Chinese Academy of
Sciences (Grant No. XDB28000000).
%\maketitle

\end{document}